\documentclass[sigconf,table]{acmart}

\copyrightyear{2026}
\acmYear{2026}
\setcopyright{cc}
\setcctype{by}
\acmConference[ICPC '26]{34th IEEE/ACM International Conference on Program Comprehension}{April 12--13, 2026}{Rio de Janeiro, Brazil}
\acmBooktitle{34th IEEE/ACM International Conference on Program Comprehension (ICPC '26), April 12--13, 2026, Rio de Janeiro, Brazil}
\acmPrice{}
\acmDOI{10.1145/3794763.3794813}
\acmISBN{979-8-4007-2482-4/2026/04}

\AtBeginDocument{%
  \providecommand\BibTeX{{%
    \normalfont B\kern-0.5em{\scshape i\kern-0.25em b}\kern-0.8em\TeX}}}

\usepackage{subcaption}

\usepackage{makecell}
\usepackage{verbatim}
\usepackage{url}
\usepackage{listings}

\usepackage[ruled,linesnumbered]{algorithm2e}

\usepackage{fontawesome}
\usepackage{bbding}

\usepackage{mathtools}
\usepackage{multirow}

\usepackage{soul}
\usepackage{enumitem}

\newcommand{\codefont}[1]{\texttt{#1}}
\newcommand{\discore}{$DIScore$}
\newcommand{\codellm}{\codefont{CodeShell-1B}}
\newcommand{\vsloss}{\codefont{$valid$-$loss$}}

\newcommand{\done}[1]{}
\newcommand{\remove}[1]{}
\newcommand{\parabf}[1]{\noindent\textbf{#1}}
\newcommand{\ptextbf}[1]{\noindent\textbf{#1}}
\newcommand{\codeIn}[1]{\texttt{#1}}
\definecolor{shadecolor}{gray}{0.92} 
\newcommand{\mybox}[1]{%
  \par\vspace{4pt}\noindent
  \colorbox{shadecolor}{%
    \parbox{\dimexpr\linewidth-2\fboxsep\relax}{%
      \emph{#1}%
    }%
  }%
  \par\vspace{4pt}%
}

\begin{document}
\title{An Empirical Study on Influence-Based Pretraining Data Selection for Code Large Language Models}


\author{Chengli Xing}
\orcid{0009-0001-0384-6613}
\authornotemark[1]
\affiliation{%
  \institution{Peking University}
  \country{China}
  \city{Beijing}
}
\email{xingchengli@stu.pku.edu.cn}

\author{Zhengran Zeng}
\orcid{0009-0009-8422-4522}
\authornote{These authors contributed equally to this research.}
\affiliation{%
  \institution{Peking University}
  \country{China}
  \city{Beijing}
}
\email{zhengranzeng@stu.pku.edu.cn}

\author{Gexiang Fang}
\orcid{0009-0008-0967-1333}
\authornotemark[1]
\affiliation{%
  \institution{Peking University}
  \country{China}
  \city{Beijing}
}
\email{2501110744@stu.pku.edu.cn}
\author{Rui Xie}
\authornotemark[2]
\orcid{0000-0002-1756-7746}
\affiliation{%
  \institution{Peking University}
  \country{China}
  \city{Beijing}
}
\email{ruixie@pku.edu.cn}
\orcid{0000-0002-9331-4716}
\author{Wei Ye}
\authornote{Rui Xie, Wei Ye, and Shikun Zhang are the corresponding authors.}
\affiliation{%
  \institution{Peking University}
  \country{China}
  \city{Beijing}
}
\email{wye@pku.edu.cn}
\author{Shikun Zhang}
\authornotemark[2]
\orcid{0000-0002-8576-2674}
\affiliation{%
  \institution{Peking University}
  \country{China}
  \city{Beijing}
}
\email{zhangsk@pku.edu.cn}

\begin{abstract}
Recent advancements in code large language models (Code-LLMs) have demonstrated remarkable capabilities in resolving programming-related tasks. Meanwhile, researchers have recognized that the quality of pre-training data is crucial for improving LLM performance. However, most of the existing research on pre-training data filtering has focused on general datasets, and little attention for programming datasets. In this paper, we aim to address this gap by exploring the effectiveness of a widely used general data filtering technique, i.e., data-influence-score filtering, within the context of programming-related datasets.
To this end, we first introduce a method for calculating data-influence-score for generative programming tasks which involves transforming a variety of downstream coding tasks into validation sets and using the model's loss on these sets as a performance metric. 
Next, we pre-train a Code-LLMs with 1 billion parameters from scratch on a dataset of 100 billion code tokens. Based on it, we conduct an extensive empirical study to evaluate the effectiveness of data-influence-score filtering methods. Specifically, we examine how well this technique improves model performance, investigate how the characteristics of beneficial training data vary across different training stages and programming tasks, and assess the feasibility of prediction-based data-influence-score filtering method.
Our findings show that data-influence-score filtering based on validation-set-loss can enhance model's programming performance. Moreover, we observe that the criteria of beneficial training data differ significantly across various downstream programming tasks. Additionally, our results suggest that predicting the oracle data-influence-score accurately is challenge.
Lastly, this study provides valuable insights into the filtering and optimization of training data for Code-LLMs, offering a foundation for future research in this domain.

\end{abstract}


\maketitle

\section{Introduction}
As Code Large Language Models (Code-LLMs) continue to evolve, they are becoming indispensable tools for enhancing productivity in software development~\cite{codesurvey,codellmsurvey,codeagentsurvey}. These Code-LLMs have shown remarkable capabilities in tasks such as code generation~\cite{codex,codereval,codeagent1,codeagent2}, bug fixing~\cite{swebench,agentless,nl2fix}, and code comprehension~\cite{coderujb,codesum1,codesum2}. Meanwhile, through extensive use and research, it has become clear that the quality of pre-training data plays a critical role in improving model performance and generalization~\cite{llmsurvey,codellmsurvey,phi1}. High-quality pre-training data not only equips models with rich and accurate programming knowledge, allowing them to better learn syntax and semantic rules, but also accelerates the training process, leading to significant improvements in accuracy and reliability for real-world programming tasks~\cite{codellmsurvey}.
Despite this recognition, there remains a gap in research on how to effectively select and optimize pre-training data for Code-LLMs. It is important to acknowledge that general text data and code data differ substantially. General text data, often sourced from news articles, books, and web content, is rich in language expressions and covers a wide range of topics and world knowledge~\cite{llmsurvey}. However, it frequently lacks the rigorous logical structure required for programming. In contrast, code data, while less abundant in world knowledge, adheres to strict syntax rules and is grounded in logical reasoning necessary for constructing functional programs~\cite{codellmsurvey}. As a result, insights derived from the study of general textual data may not directly apply to code data, emphasizing the need for specialized research in this area~\cite{influence2,mates,dsdm}.
With this in mind, it is crucial to conduct in-depth investigations into the quality of pre-training code data and to develop robust data filtering and scheduling strategies. Such efforts are essential not only for optimizing the performance of Code-LLMs but also for expanding their applicability across a broader range of programming tasks and environments.

Recent research~\cite{santacoder,starcoder,ppl,phi1,askllm,qurating,mates,dsdm} in the pre-training of LLMs has demonstrated the effectiveness of several data filtering techniques, including data deduplication~\cite{santacoder,starcoder}, perplexity-based filtering~\cite{ppl}, and LLM-score-based filtering~\cite{phi1,askllm,qurating}. Furthermore, emerging studies have highlighted the promise of data-influence-score (denoted as \discore{} in subsequent sections) filtering methods. Unlike traditional approaches that focus solely on the inherent characteristics of training data~\cite{bigscience,ppl,phi1}, \discore{} filtering methods consider the impact of individual data samples on downstream task performance. Specifically, this approach involves performing a single training step with a given sample and then evaluating the performance improvement on selected downstream tasks before and after this single training step. This improvement serves as a quality score for the sample in data filtering~\cite{mates,dsdm}.
Building on the success of \discore{} filtering in general text data, our research aims to explore its potential in filtering code pre-training data (e.g., source code from GitHub). First, we will assess the effectiveness of the \discore{} method in identifying valuable code data and examine how the importance of specific training data may vary at different stages of model training (i.e., different checkpoints). Additionally, since different programming tasks (e.g., Python vs. SQL) require distinct capabilities from the model, we will investigate the commonalities and differences in valuable training data across various programming domains.
Next, we will compare the results of the data influence filtering method with other commonly used techniques, such as perplexity filtering and LLM score-based filtering, to better understand their relative strengths and weaknesses. Finally, we will evaluate the effectiveness of current \discore{} filtering schemes, particularly those based on \discore{} prediction (e.g., predicts the \discore{} base on another small model).

To achieve this, we need an effective way to measure the impact of pre-training data on the performance (i.e, \discore{}) of software engineering tasks. Previous studies~\cite{mates,dsdm} have primarily focused on classification tasks, using the improvement of downstream tasks $accuracy$ as the \discore{}. However, most code-related tasks are generative tasks, and their evaluation has increasingly shifted towards execution-based methods~\cite{codex,coderujb}. While execution-based evaluation can be more reflective of real-world performance, it is computationally expensive (e.g., pre-training data filtering would require executing billions of programs) and often fails to capture the fine-grained influence of individual training samples. This limitation makes it challenging to directly apply existing \discore{} scoring methods to code tasks.
Therefore, we propose transforming various downstream code tasks into validation sets and using model $loss$ on these sets as an efficient proxy for evaluating performance. By leveraging $loss$ as the evaluation metric, we can more precisely gauge the impact of individual training samples on the model’s performance across different code-related tasks. This approach not only provides a more scalable solution, but also offers a finer-grained understanding of how specific data influences the model's ability to generalize and perform effectively in diverse programming scenarios.

Next, we trained a CodeLLM with 1 billion (B) parameters from scratch (denoted as \codeIn{CodeLLM-1B} in subsequent sections), using a pre-training dataset consisting of 100 billion (B) code tokens. \codeIn{CodeLLM-1B} served as the foundation for a comprehensive empirical investigation aimed at addressing our key research questions in detail.
First, we validated the effectiveness of our evaluation framework, which is based on the validation-set-loss (denoted as \vsloss{} in subsequent sections) and the \discore{} filtering method. Our findings indicate that \discore{} (measured by \vsloss{}) filtering can significantly improve the model's performance on practical evaluation metrics (e.g., $pass@k$, $accuracy$, or $Exact Match$).
Subsequently, we examined how the \discore{} of pre-training data varies across different stages of training and across different programming tasks. Our analysis revealed that as training progresses, the criteria for identifying beneficial samples evolve and eventually stabilize during the later stages of training. Additionally, we found that different downstream programming tasks have distinct criteria for what constitutes beneficial training data. This highlights the importance of constructing validation sets that are both general and practical, ensuring they meet the diverse requirements of various tasks.
We then explored the similarities and differences between traditional data filtering methods and the \discore{} filtering approach. The results showed that the samples identified as beneficial by perplexity-based~\cite{ppl} and LLM-score-based~\cite{phi1,askllm,qurating} filtering methods differ significantly from those selected by the \discore{} filtering method. This suggests that conventional filtering methods may not be as effective in identifying the training data that is most beneficial for enhancing model performance.
Finally, we assessed the effectiveness of using smaller models to predict \discore{}. Our findings revealed that strategies relying on small models to score data influence often struggle to accurately predict which training samples are beneficial. This leads to the selection of lower-quality data, suggesting that using small models to approximate the true data influence distribution presents significant challenges, thereby limiting the immediate practical applicability of this method for large-scale pre-training.

In conclusion, the key contributions of this study are as follows:

\begin{itemize}
\item \textbf{Technique}: We employed an efficient method for applying \discore{} filtering to programming data. By transforming various downstream code tasks into validation sets and using the model's loss (\vsloss{}) on these sets as a performance proxy, we accurately measure the influence of individual data samples. This approach overcomes the limitations of traditional \discore{} measurement in generative code tasks and offers a practical solution for pre-training data filtering.

\item \textbf{Empirical Study}: We trained a 1B parameter CodeLLM from scratch on a 100B token dataset. This model allowed us to conduct an in-depth analysis of several key research questions, where we (1) verified the effectiveness of our \discore{} filtering method, (2) investigated how data influence evolves across training stages and tasks, (3) compared our approach with mainstream filtering methods, and (4) evaluated the efficacy of using smaller models for \discore{} prediction.

\item \textbf{Findings and Insights}: Our research yielded several important findings for optimizing code pre-training data selection. Specifically, we demonstrate that (1) \vsloss{}-based \discore{} filtering is effective at enhancing model performance. (2) The criteria for beneficial data evolve during training and stabilize in later stages, suggesting that filtering strategies should be dynamic. (3) Different downstream tasks have varying standards for useful data, highlighting the need for general and practical validation sets. (4) Using smaller models to predict \discore{} yields low accuracy and selects suboptimal data, indicating this approach requires further optimization.
\end{itemize}

Ultimately, our goal is to provide a comprehensive evaluation framework that assesses the impact of training data, thereby offering guidelines for future data selection strategies and improving model performance.

\section{Background and Related Work}
\subsection{Code Large Language Models}
Recently, Code-LLMs have made remarkable progress in addressing various programming tasks. These Code-LLMs are typically built upon a Transformer-Decoder~\cite{transformer} architecture and are pre-trained on vast amounts of code datasets, enabling them to generate high-quality code snippets, auto-completing code, and even fixing bugs or performing code repairs. Prominent models like OpenAI's Codex~\cite{codex}, and open-source alternatives such as CodeLlama~\cite{codellama}, StarCoder~\cite{starcoder,starcoder2}, Qwen-Coder~\cite{qwen2}, and DeepSeek-Coder~\cite{deepseekcoder} have been successfully applied in software development.

Compared to general-purpose LLMs, Code-LLMs exhibit distinct advantages due to their specialized training. This specialization grants them a deeper understanding of code syntax and semantics, enabling them to generate logically coherent and syntactically correct outputs. Furthermore, the inherent logical structure of code data helps these models develop strong reasoning abilities, making them adept at tasks requiring complex algorithms and problem-solving~\cite{codellmsurvey,codex}.

\subsection{Pre-training Data Filtering}
High-quality pre-training data is crucial for LLM performance, making data filtering a vital step to curate relevant, diverse, and high-quality datasets. Prevailing filtering strategies are predominantly rule-based, employing step-by-step procedures~\cite{llmsurvey} like deduplication and word-frequency filtering~\cite{santacoder}. While effective for quickly removing noise from large datasets, these methods often fail to address more nuanced aspects of data quality, such as semantic coherence or structural integrity~\cite{askllm,ppl,mates}.

To overcome these limitations, some studies have adopted perplexity-based filtering~\cite{ppl0,ppl}. Perplexity measures how well a model predicts a data sample; a lower score indicates a better fit with the model's learned distribution~\cite{ppl}. By filtering out high-perplexity samples, researchers can improve dataset quality. For example, Raffel et al.~\cite{ppl0} used this method to construct the C4 dataset by excluding low-quality web text. Recent works have also explored optimizing data mixtures by predicting selection efficiency~\cite{DataMixing}.

Beyond perplexity, other research has explored using LLMs themselves for data scoring~\cite{phi1,qurating,askllm}. A notable example is the Phi series~\cite{phi1}, which employed LLMs to rate data based on its "educational value." This approach helped identify well-structured and thoroughly annotated code, allowing the model to learn more effectively from high-quality examples.

In summary, the evolution of data filtering, i.e., from simple rules to sophisticated perplexity and LLM-based scoring, has been crucial in enhancing pre-training dataset quality. These advancements ensure models train not only on clean data but also on data better aligned with their target tasks, ultimately boosting performance.

\subsection{Data Influence}
However, existing filtering methods typically focus on static data properties (e.g., syntax, frequency), overlooking the dynamic impact of data during training. To address this, some researchers have argued that it is crucial to assess the actual influence of data on model performance (referred to as data influence~\cite{influence}) as a more fundamental criterion for data filtering~\cite{less,dsdm,mates}. By measuring this contribution, we can prioritize data that positively impacts model outcomes.

Data influence has been previously explored in various contexts, such as identifying mislabeled samples~\cite{mislabeled}, analyzing model memorization~\cite{memorize}, and enhancing interpretability~\cite{interpretability}. For LLMs, however, the prohibitive computational cost of calculating influence functions has limited their practical application at scale~\cite{dsdm,mates}.

Despite these challenges, several methods have attempted to integrate data influence into data selection. For example, LESS~\cite{less} and MATES~\cite{mates} applied influence functions during fine-tuning and pre-training, respectively. Yet, the vast scale of pre-training data presents a significant hurdle. To mitigate this, MATES used a proxy model to approximate influence scores, but its accuracy was constrained by the proxy's limited capacity~\cite{mates}. Furthermore, an over-reliance on influence functions can risk reducing data diversity, potentially harming model generalization.

Crucially, the application of data influence filtering to code data remains largely unexplored~\cite{influence2,mates,dsdm}. Most existing studies focus on classification tasks, which differ significantly from the generative nature of code-related tasks. This leaves open the question of how to effectively apply and evaluate data influence methods for code pre-training datasets. This study aims to bridge this gap by systematically investigating data influence filtering for code pre-training. We explore its effectiveness at various training stages and across different programming tasks, offering insights into optimizing data selection for Code-LLMs. Ultimately, our goal is to provide a framework that enhances both the quality and diversity of training data, thereby improving model performance.

\section{The Empirical Study}
\subsection{Research Questions}
\label{sec:rq}
\ptextbf{RQ1: How effective are the validation-set-loss evaluation method and the data-influence-score filtering method?}

Before utilizing \vsloss{} to calculate \discore{} (detailed in Section~\ref{sec:subject3}), we first need to validate its effectiveness to ensure its relevance to real-world task metrics. To do so, we pre-train a 1B parameter model on 100B code tokens and examine the correlation between its \vsloss{} and standard evaluation metrics (e.g., $pass@k$, $accuracy$, $BLEU$) across various checkpoints and tasks. Next, we use the validated metric to calculate \discore{} for each data point and continue pre-training with high \discore{} data to assess the effectiveness of this filtering strategy.

\ptextbf{RQ2: How do data-influence-scores vary across different pre-training stages and programming tasks?}

We then analyze how the \discore{} distribution for a given training set changes across pre-training stages (i.e., different checkpoints) and downstream tasks. This investigation aims to determine whether data quality is static or dynamic relative to the model's training progress. Understanding this variability is crucial for assessing the adequacy of static filtering methods (e.g., perplexity) versus dynamic, curriculum-learning-style data selection strategies.

\ptextbf{RQ3: How do mainstream data filtering methods compare with the data-influence-score filtering method?}

This research question explores \discore{} filtering method against mainstream perplexity and LLM-scoring methods. The goal is to determine if these existing approaches can implicitly identify high \discore{} data, thereby revealing their relative strengths, weaknesses, and overlaps with our influence-based approach.

\ptextbf{RQ4: How effective is the prediction-based data-influence-score filtering method?}

This research question investigates the feasibility of using a smaller, cost-effective model to predict \discore{}. We train this proxy model to estimate influence, filter the training data based on these predictions, and evaluate the effectiveness of this prediction-based filtering strategy on subsequent model training.

\subsection{Study Subjects}
\label{sec:subject}
This subsection provides a detailed overview of the sources and composition of the subjects involved in the study.

\subsubsection{CodeLLM-1B}
\label{sec:subject1}
\begin{sloppypar}
To investigate data filtering strategies for code data in detail, we trained a 1B parameter model, named \codellm{}, from scratch. The model was trained on 100B tokens of code data and is based on the widely adopted CodeLlama~\cite{codellama} architecture. By referencing the model structure parameters of TinyLlama~\cite{tinyllama}, we adjusted the model size to 1.1B parameters. For tokenization, we used the pre-existing tokenizer from CodeLlama. 
As for the pre-training data, we utilized the open-source StarCoderData~\cite{starcoder} dataset, which is commonly used for code pre-training. Then, we randomly selected 100B tokens from the total 260B tokens available in StarCoderData. Training from scratch, rather than continue pre-training~\cite{codellama} an existing model like StarCoder, was a deliberate choice to ensure a controlled experimental environment. This approach allows us to isolate the effects of our data filtering strategies by eliminating confounding variables from prior training stages, which is a common practice in this line of pre-training data filtering research~\cite{askllm,ppl,mates,dsdm}.
\end{sloppypar}

For the learning rate strategy, we followed prior work~\cite{mates} and adopted the Warmup-Stable-Decay (WSD)~\cite{minicpm} learning rate schedule. This approach ensures that the learning rate remains stable at $1e^{-4}$ for the majority of the training process, allowing for a more consistent and fair comparison of \discore{} distribution across different training stages. Additionally, all other training hyperparameters were kept consistent with previous studies~\cite{mates}.
During training, we saved a checkpoint every 5B tokens processed. These checkpoints enabled us to evaluate the performance trends of \codellm{} on downstream programming tasks and to investigate how the \discore{} distribution changes over different training stages. The entire pre-training process required 560 GPU hours on A100-80GB GPUs.

\subsubsection{Evaluation Datasets}
\label{sec:subject2}

\begin{table}[]
\caption{Statistics of evaluation datasets.}
\label{tab:sed}
\resizebox{\columnwidth}{!}{
\begin{tabular}{|c|c|c|c|c|c|}
\hline
\textbf{Datasets} & \textbf{Language} & \textbf{Size} & \textbf{Execution} & \textbf{Metric} & \textbf{With GT} \\ \hline\hline
Humaneval~\cite{codex}         & Python, Java, C++   & 164           & \checkmark                & $Pass@1$          & $\times$                         \\ \hline
MBPP~\cite{mbpp}              & Python            & 500           & \checkmark                & $Pass@1$          & \checkmark                        \\ \hline
DS-1000~\cite{DS1000}           & Python            & 1000          & \checkmark                & $Accuracy$        & \checkmark                        \\ \hline
CrossCodeEval~\cite{crosscodeeval}     & Python            & 2665          & $\times$                 & $ExactMatch$     & \checkmark                        \\ \hline
Bird-SQL~\cite{birdsql}          & SQL               & 1533          & \checkmark                & $Accuracy$        & \checkmark                        \\ \hline
\end{tabular}
}
\end{table}

To comprehensively evaluate \codellm{}'s programming capabilities, we selected multiple downstream task datasets, as detailed in Table~\ref{tab:sed}. Our study incorporates five widely used benchmarks spanning four major languages, including foundational datasets like \textit{HumanEval} and \textit{MBPP}, the more challenging \textit{CrossCodeEval}, the data science-focused \textit{DS-1000}, and the SQL-centric \textit{Bird-SQL}. In Table~\ref{tab:sed}, "Size" refers to the number of problems, and "Execution" denotes the evaluation method. All benchmarks except \textit{CrossCodeEval}, which uses $ExactMatch$~\cite{crosscodeeval}, rely on execution-based metrics. The "With GT (Ground-Truth)" column shows that all datasets provide reference answers except for \textit{HumanEval}.

Furthermore, to study the \discore{} of training data, we constructed a validation set from a subset of these datasets. For benchmarks with provided ground-truth (GT) answers, we directly used their problems and reference solutions. For the \textit{HumanEval} series, which lacks GT answers, we generated reference solutions using CodeLlama-34B~\cite{codellama}. Meanwhile, we recognized that even a powerful model like CodeLlama-34B cannot guarantee correctness for every problem. Consequently, we excluded particularly complex problems that were deemed unsuitable for evaluating a 1B model. Finally, to ensure a consistent validation size across tasks and reduce computational overhead, we standardized the validation sets. For datasets with over 200 problems, we randomly sampled 200 to form the final validation set, while for those with fewer than 200 problems (e.g., \textit{HumanEval}), we utilized the entire dataset for our analysis.

\subsubsection{Data Influence}
\label{sec:subject3}
The calculation of \discore{} is a central component of this research. In prior studies~\cite{influence,dsdm,mates}, \discore{} has been defined as the improvement in performance observed on downstream tasks before and after training the model with a single data point. We formalize this calculation in Equation~\ref{eq:influence}~\cite{mates}:

\begin{equation}
I_M(x_i; D_r) = L(D_r\ |\ M) - L(D_r\ |\ A(M, x_i))
\label{eq:influence}
\end{equation}

In this equation, $M$ represents the current state of the model, $x_i$ is the data point for which we wish to calculate the \discore{}, and $D_r$ is the validation set associated with the downstream task. Thus, $I_M(x_i; D_r)$ quantifies the \discore{} of the data point $x_i$ by measuring the change in $loss$ on the validation set $D_r$, before and after updating the model $M$ with a single training step on $x_i$. Specifically, it is the difference between the $loss$ of the current model $M$ (i.e., $L(D_r\ | \ M)$) and the $loss$ of the updated model $A(M, x_i)$ (i.e., $L(D_r\ | \ A(M, x_i))$), where $A(M, x_i)$ denotes the model after being trained on $x_i$.

It should be noted that prior studies~\cite{dsdm,mates} have predominantly focused on \discore{} of classification tasks, where $loss$ values are closely aligned with downstream metrics (e.g., $accuracy$). In such cases, changes in $loss$ values (i.e., \discore{}) tend to correlate well with changes in task performance. However, in the domain of software engineering, many programming tasks are generative. The evaluation criteria for these tasks have shifted from similarity-based metrics to execution-based correctness measures. 
This shift presents a challenge that \discore{} calculated from \vsloss{} may not fully capture the actual trends in downstream performance metrics for generative tasks (e.g., $pass@k$). Therefore, this study aims to investigate the effectiveness of \vsloss{} based \discore{} calculations specifically within the context of generative programming tasks.

\subsection{Results and Analysis}

\subsubsection{RQ1: How effective are the validation-set-loss evaluation method and the data-influence-score filtering method?\\}
\label{sec:rq1}

This research question investigates the effectiveness of using \vsloss{} and \discore{} filtering methods. To do so, we trained a \codellm{} on 100B code tokens, saving 20 checkpoints at 5B-token intervals to monitor performance progression (detailed description of the training process can be found in Section~\ref{sec:subject1}). We then evaluated these checkpoints on various downstream programming tasks using corresponding validation sets, and the details processes are provided in Section~\ref{sec:subject2}.

\begin{figure}[ht]
  \centering
  \includegraphics[width=1\linewidth]{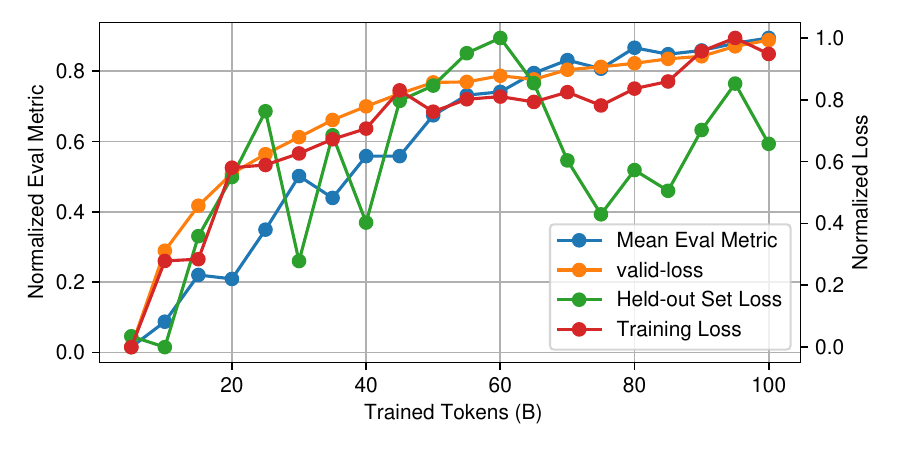}
  \caption{Comparison of actual performance metrics and different loss-based metrics over training checkpoints.}
  \label{fig:comparison}
\end{figure}

\begin{table*}[ht]
\caption{Spearman rank correlation (p-values) between actual performance metrics and different loss-based metrics across different programming tasks}
\label{tab:spearmanr}

\resizebox{\linewidth}{!}{
\begin{tabular}{|c|c|c|c|c|c|c|c|}
\hline
\textbf{Loss Metrics} & \textbf{HumanEval-Py} & \textbf{HumanEval-Java} & \textbf{HumanEval-Cpp} & \textbf{MBPP}   & \textbf{CrossCodeEval} & \textbf{Bird-SQL} & \textbf{DS-1000} \\ \hline\hline
\textbf{valid-loss}    & 0.9578 (3.3e-11)      & 0.8465 (2.5e-06)        & 0.8842 (2.3e-07)       & 0.9233 (6.5e-9) & 0.9612 (1.6e-11)       & 0.7368 (2.1e-4)   & 0.8421 (3.2e-6)  \\ \hline
\textbf{Training Loss} & 0.9007 (6.1e-08)      & 0.7863 (3.9e-05)        & 0.8601 (1.1e-6)        & 0.9112 (2.3e-8) & 0.8882 (1.7e-7)        & 0.7187 (3.5e-4)   & 0.7984 (2.4e-5)  \\ \hline
\textbf{Held-out Loss} & 0.3984 (8.1e-2)       & 0.5379 (1.4e-2)         & 0.4721 (3.5e-2)        & 0.5022 (2.4e-2) & 0.4968 (2.5e-2)        & 0.2526 (2.8e-1)   & 0.4497 (4.6e-2)  \\ \hline
\end{tabular}
}
\end{table*}

Once the validation sets were established, we computed the \vsloss{} for each of the 20 checkpoints. Figure~\ref{fig:comparison} illustrates the average normalized (i.e., the results for each task are linearly mapped to the $[0,1]$ range) actual performance (i.e., $pass@1$, $accuracy$ and $ExactMatch$) trends of \codellm{} on the 7 downstream tasks alongside the corresponding normalized \vsloss{} trends over the course of training. Additionally, we include the normalized training loss and normalized held-out set loss (i.e., keep some training data out of training, just for validation) for comparison, as these are commonly used performance monitoring metrics. As shown, both the model’s performance on downstream tasks and the \vsloss{} improved significantly as training progressed. While there were some fluctuations, the overall trends of both metrics remained aligned. Importantly, the \vsloss{} demonstrated a much stronger alignment with actual performance metrics compared to training loss and held-out set loss. This suggests that the \vsloss{} provides a more accurate reflection of the model’s real-world performance on downstream tasks, reinforcing its reliability as an evaluation method.

To quantify this relationship, we computed the Spearman rank correlation~\cite{spearman} between actual performance and the loss metrics for each task. The results in Table~\ref{tab:spearmanr} reveal a strong, statistically significant positive correlation between \vsloss{} and performance across all tasks. For instance, even the lowest correlation for \textit{BirdSQL} is high at $0.7368$ (p-value = $2.1e$-$4$). Furthermore, \vsloss{} consistently shows a higher correlation than both training loss and held-out loss.

These findings confirm that \vsloss{} is a highly reliable proxy for actual downstream task performance. Its strong correlation and computational efficiency make it an excellent alternative to running full, resource-intensive evaluations.

\mybox{Finding 1: The \vsloss{} trends closely align with the trends of actual evaluation metrics and outperform commonly used performance monitoring metrics. Therefore, the \vsloss{} is an efficient and reliable alternative for downstream task evaluation.}

\begin{figure*}[htb]
\centering
\includegraphics[width=1\linewidth]{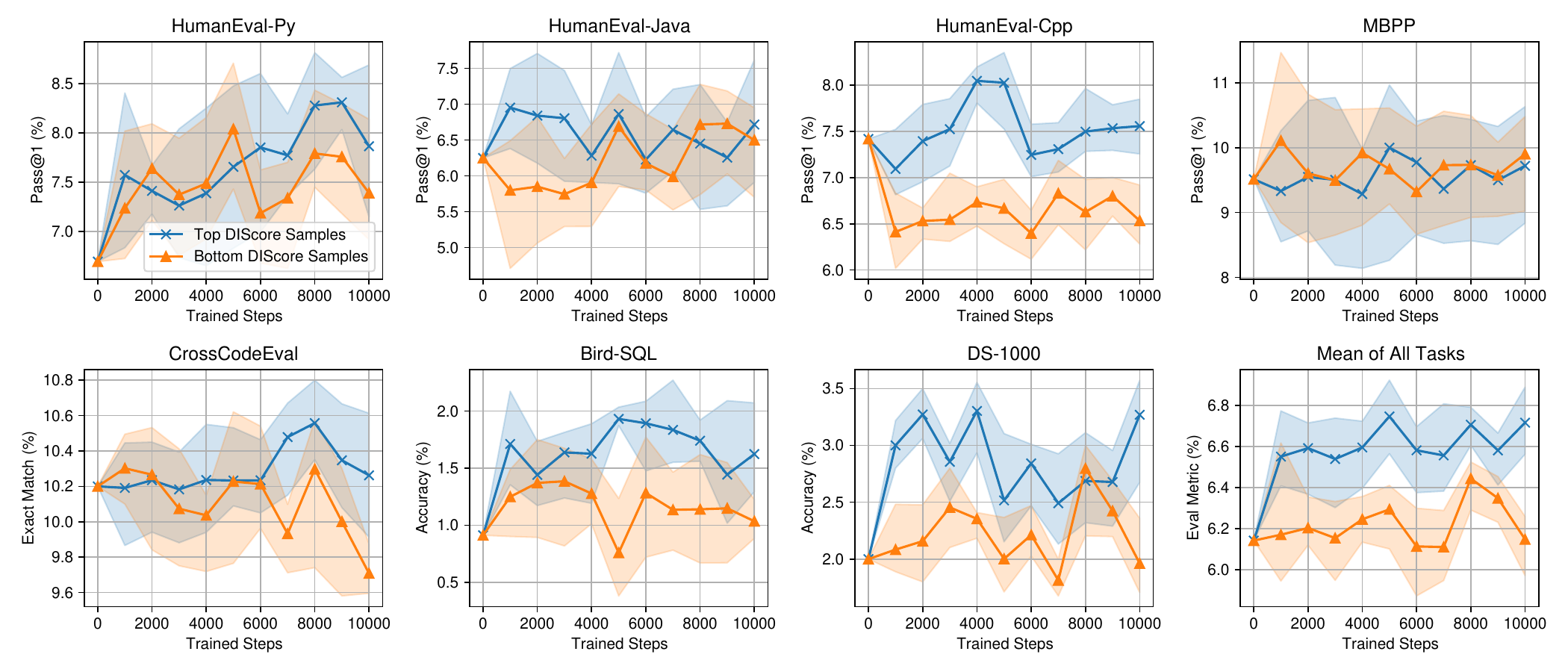}
\caption{Performance trends on downstream programming tasks for models trained with Top and Bottom Samples selected by the \discore{}.}
\label{fig:selecttrain}
\end{figure*}

After confirming the effectiveness of the \vsloss{}, we further use the change in \vsloss{} after a single-step training on individual samples as a metric to measure \discore{}. A detailed explanation of this metric and its calculation process is provided in Section~\ref{sec:subject3}.
Next, we constructed a small, multilingual training dataset consisting of code from 10 different programming languages. Our goal was to examine how training data from various programming languages impacts performance on different downstream programming tasks. Of these 10 languages, 7 are widely used, while the remaining 3 are less common. This setup allowed us to compare the influence of mainstream versus rare programming languages on the tasks. Specifically, we randomly selected 20,000 training samples in total from the remaining 160B tokens of the StarCoderData~\cite{starcoder} dataset (i.e., data not seen by \codellm{}\codefont{-CP100B} during its initial pre-training), with 2,000 samples per language. We then performed single-step training on the \codellm{}\codefont{-CP100B} (i.e., \codellm{} trained on 100B code tokens) to compute the \discore{} of each training samples.
Moreover, by accurately measuring the \discore{} for the validation sets of various programming tasks, we can determine the \discore{} for each individual task. By averaging these values across all the tasks, we obtain the overall \discore{}.

Once the \discore{} were computed, we ranked the 20,000 training samples in descending order of influence for each programming task. For each task, we selected the top 10,000 samples with the highest \discore{}, labeling them as "Top \discore{} Samples" (i.e., considered beneficial by the \discore{}). Conversely, we labeled the bottom 10,000 samples with the lowest \discore{} as "Bottom \discore{} Samples" (i.e., considered harmful by the \discore{}).
To evaluate the effectiveness of this \discore{} filtering method, we continued pre-training the \codellm{}\codefont{-CP100B} using the selected "Top and Bottom \discore{} Samples". The training hyperparameters were kept consistent with those used during the initial pre-training phase.

To ensure the reliability of our results, we conducted 5 independent experiments, each repeated 5 times, covering every step from \discore{} calculation to evaluation. Figure~\ref{fig:selecttrain} presents the mean performance, along with the minimum and maximum values, across these 5 experiments for 7 downstream programming tasks. Note that the "Mean" performance is calculated by averaging the normalized scores of all tasks.
The results show that \codellm{} trained with the "Top \discore{} Samples" generally outperforms the model trained with the "Bottom \discore{} Samples" across most tasks. This demonstrates that the \discore{} filtering method, based on the \vsloss{}, is effective in identifying and selecting data that is beneficial for downstream tasks.

\mybox{Finding 2: The \discore{} filtering method, grounded in the \vsloss{}, is effective at selecting data that is beneficial for downstream programming tasks.}

\subsubsection{RQ2: How do data-influence-scores vary across different pre-training stages and programming tasks?\\}

In RQ1, we found that \discore{} filtering successfully identifies beneficial data at a given checkpoint. In this research question, we investigate the consistency of this beneficial data across different training stages and downstream tasks. 
To begin, we examine how data influence evolves over time by selecting five checkpoints from the \codellm{} training process (at 20B, 40B, 60B, 80B, and 100B tokens). Using these checkpoints, we calculated the \discore{} for the 20,000 samples from Section~\ref{sec:rq1} across 7 different programming task validation sets.

\begin{figure}
  \centering
  \includegraphics[width=0.6\linewidth]{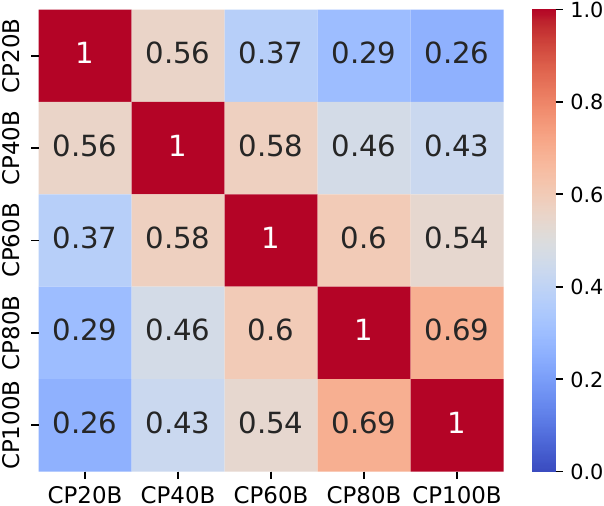}
  \caption{Spearman rank correlation of \discore{} between different training stages of \codellm{}.}
  \label{fig:spearmancc}
\end{figure}

Next, we computed the Spearman correlation coefficients for the \discore{} between each pair of these five checkpoints. The results, shown in Figure~\ref{fig:spearmancc}, reveal that the influence of training data evolves as the model matures. For instance, the correlation between the 20B and 40B token checkpoints is 0.56, but this value drops progressively to 0.37, 0.29, and 0.26 when comparing the 20B checkpoint to the 60B, 80B, and 100B checkpoints, respectively. This indicates that the definition of "beneficial data" shifts significantly during training; the further apart the stages, the greater the divergence. This observation suggests that static filtering methods, such as Perplexity and LLM-scoring, are likely insufficient for identifying optimal data throughout the entire training process.

Conversely, we also observed that the set of beneficial samples tends to stabilize in the later stages of training. The correlation between the 60B and 80B checkpoints increases to 0.60, and it rises further to 0.69 between the 80B and 100B checkpoints. This suggests that as the model's parameters converge, its assessment of data influence becomes more consistent.

\mybox{Finding 3: The selection of beneficial training data varies across different training stages. While the beneficial data between adjacent checkpoints is relatively similar, the criteria for what is considered beneficial evolves as the model's parameters are updated. However, in the later stages of training, as the model's parameters stabilize, the model's perception of beneficial data becomes more consistent.}

\begin{figure}
\centering
\includegraphics[width=1\linewidth]{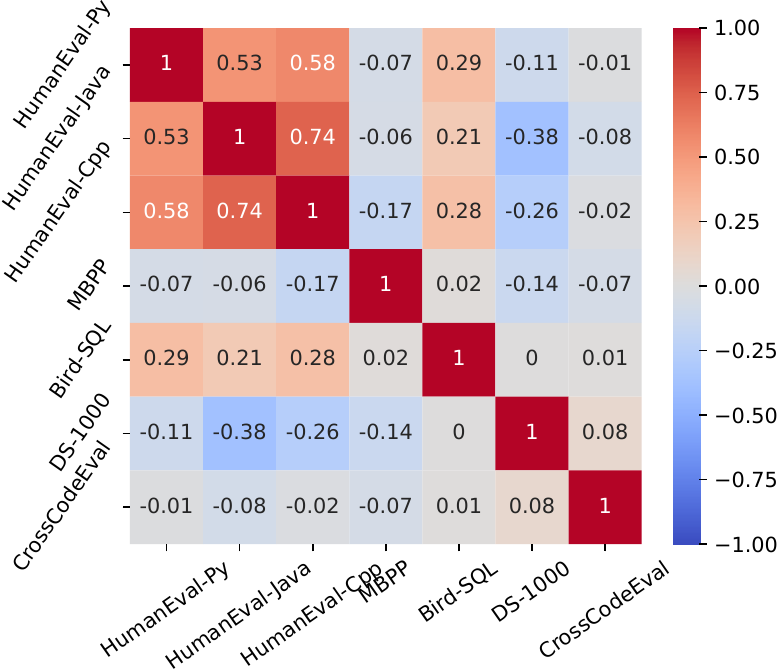}
\caption{Spearman correlation coefficients of \discore{} between different programming tasks based on the \codellm{}\codefont{-CP100B}.}
\label{fig:spearman-task}
\end{figure}

Next, we explored how \discore{} varies across different programming tasks. Figure~\ref{fig:spearman-task} presents the Spearman correlation coefficients of \codellm{}\codefont{-CP100B} across seven downstream tasks. The results reveal that certain tasks exhibit high correlations; for instance, all tasks within the \textit{HumanEval} series have correlations exceeding 0.50. Additionally, the SQL-based \textit{BirdSQL} task shows a moderate correlation with the \textit{HumanEval} series (coefficients > 0.20).

Conversely, other tasks like \textit{MBPP}, \textit{DS-1000}, and \textit{CrossCodeEval} demonstrate no significant correlation with the others. Notably, \textit{DS-1000} even displays negative correlations with the \textit{HumanEval} series (coefficients of -0.11, -0.38, and -0.26). This suggests that the same training data can have drastically different, even opposing, influences depending on the coding environment or task scenario.

This analysis also highlights that task content similarity is a key factor in data influence. The \textit{HumanEval} series, which involves the same task across different languages, shows strong correlation. In contrast, sharing a programming language alone does not guarantee correlated influence. For example, despite all being Python-based, \textit{HumanEval-Python}, \textit{MBPP}, \textit{DS-1000}, and \textit{CrossCodeEval} exhibit no notable correlation in their \discore{}.

\mybox{Finding 4: The influence of training data is highly task-dependent, varying significantly across different coding scenarios. While tasks with similar content (e.g., the HumanEval series) show correlated influence, most exhibit little to no relationship, with some even showing negative correlations.}

\begin{figure}
\centering
\includegraphics[width=1\linewidth]{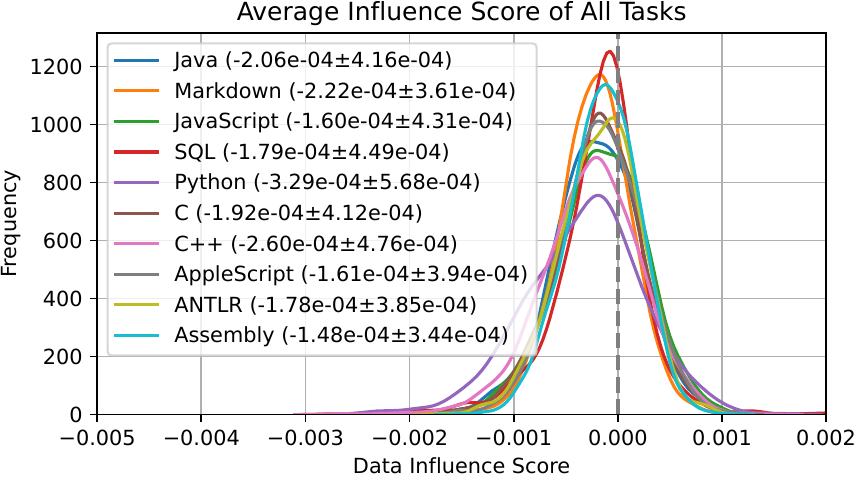}
\caption{Distribution of \discore{} on different programming languages.}
\label{fig:language}
\end{figure}

\begin{table}[]
\centering
\caption{Statistical Summary of \discore{} Distributions (Mean and Standard Deviation) Across Tasks.}
\label{tab:language}
\resizebox{\linewidth}{!}{%
\begin{tabular}{|c|cccc|}
\hline
\textbf{Language} & \textbf{\begin{tabular}[c]{@{}c@{}}HumanEval\\-Py\end{tabular}} & \textbf{\begin{tabular}[c]{@{}c@{}}HumanEval\\-Cpp\end{tabular}} & \textbf{Bird-SQL} & \textbf{DS-1000} \\ \hline\hline
\textbf{Java} & \begin{tabular}[c]{@{}c@{}}-1.43e-04\\ (6.78e-04)\end{tabular} & \begin{tabular}[c]{@{}c@{}}-5.87e-05\\ (6.75e-04)\end{tabular} & \begin{tabular}[c]{@{}c@{}}-7.32e-05\\ (1.08e-03)\end{tabular} & \begin{tabular}[c]{@{}c@{}}-3.16e-04\\ (1.44e-03)\end{tabular} \\ \hline
\textbf{Markdown} & \begin{tabular}[c]{@{}c@{}}-1.52e-04\\ (5.15e-04)\end{tabular} & \begin{tabular}[c]{@{}c@{}}-6.70e-05\\ (4.30e-04)\end{tabular} & \begin{tabular}[c]{@{}c@{}}-1.40e-04\\ (1.09e-03)\end{tabular} & \begin{tabular}[c]{@{}c@{}}-2.47e-04\\ (1.43e-03)\end{tabular} \\ \hline
\textbf{JavaScript} & \begin{tabular}[c]{@{}c@{}}-2.31e-05\\ (6.80e-04)\end{tabular} & \begin{tabular}[c]{@{}c@{}}-1.47e-05\\ (6.52e-04)\end{tabular} & \begin{tabular}[c]{@{}c@{}}9.63e-06\\ (1.06e-03)\end{tabular} & \begin{tabular}[c]{@{}c@{}}-3.58e-04\\ (1.62e-03)\end{tabular} \\ \hline
\textbf{SQL} & \begin{tabular}[c]{@{}c@{}}-4.86e-05\\ (4.14e-04)\end{tabular} & \begin{tabular}[c]{@{}c@{}}-4.11e-05\\ (3.61e-04)\end{tabular} & \begin{tabular}[c]{@{}c@{}}-3.14e-04\\ (2.36e-03)\end{tabular} & \begin{tabular}[c]{@{}c@{}}-3.88e-04\\ (1.20e-03)\end{tabular} \\ \hline
\textbf{Python} & \begin{tabular}[c]{@{}c@{}}-2.78e-04\\ (1.04e-03)\end{tabular} & \begin{tabular}[c]{@{}c@{}}-6.91e-05\\ (7.10e-04)\end{tabular} & \begin{tabular}[c]{@{}c@{}}5.21e-05\\ (1.16e-03)\end{tabular} & \begin{tabular}[c]{@{}c@{}}-2.54e-04\\ (2.28e-03)\end{tabular} \\ \hline
\textbf{C} & \begin{tabular}[c]{@{}c@{}}-1.08e-04\\ (6.17e-04)\end{tabular} & \begin{tabular}[c]{@{}c@{}}-1.25e-04\\ (7.17e-04)\end{tabular} & \begin{tabular}[c]{@{}c@{}}-6.62e-05\\ (1.07e-03)\end{tabular} & \begin{tabular}[c]{@{}c@{}}-3.31e-04\\ (1.37e-03)\end{tabular} \\ \hline
\textbf{C++} & \begin{tabular}[c]{@{}c@{}}-1.68e-04\\ (7.73e-04)\end{tabular} & \begin{tabular}[c]{@{}c@{}}-2.62e-04\\ (1.01e-03)\end{tabular} & \begin{tabular}[c]{@{}c@{}}2.35e-06\\ (1.12e-03)\end{tabular} & \begin{tabular}[c]{@{}c@{}}-2.88e-04\\ (1.54e-03)\end{tabular} \\ \hline
\textbf{AppleScript} & \begin{tabular}[c]{@{}c@{}}3.46e-05\\ (5.80e-04)\end{tabular} & \begin{tabular}[c]{@{}c@{}}-6.00e-05\\ (4.71e-04)\end{tabular} & \begin{tabular}[c]{@{}c@{}}-6.63e-05\\ (1.29e-03)\end{tabular} & \begin{tabular}[c]{@{}c@{}}-2.08e-04\\ (1.47e-03)\end{tabular} \\ \hline
\textbf{ANTLR} & \begin{tabular}[c]{@{}c@{}}-1.02e-06\\ 4.96e-04\end{tabular} & \begin{tabular}[c]{@{}c@{}}1.45e-05\\ (4.12e-04)\end{tabular} & \begin{tabular}[c]{@{}c@{}}-1.80e-04\\ (1.39e-03)\end{tabular} & \begin{tabular}[c]{@{}c@{}}-2.27e-04\\ (1.09e-03)\end{tabular} \\ \hline
\textbf{Assembly} & \begin{tabular}[c]{@{}c@{}}-6.54e-05\\ (5.22e-04)\end{tabular} & \begin{tabular}[c]{@{}c@{}}-7.43e-05\\ (4.44e-04)\end{tabular} & \begin{tabular}[c]{@{}c@{}}-1.56e-04\\ (1.04e-03)\end{tabular} & \begin{tabular}[c]{@{}c@{}}-1.43e-04\\ (1.17e-03)\end{tabular} \\ \hline
\end{tabular}%
}
\end{table}

We further examined how \discore{} vary across different programming languages. As illustrated in Figure~\ref{fig:language}, the average \discore{} distribution across all tasks generally follows a normal distribution centered around 0. For a more detailed breakdown, Table~\ref{tab:language} presents the specific mean and standard deviation of the \discore{} distribution for several key programming tasks. In this context, a positive \discore{} indicates that the training data is beneficial to the task.
Interestingly, the results show that even training data in the same language as the downstream task exhibits a mean \discore{} close to 0, indicating no significant average advantage over data from other languages. For instance, in the \textit{HumanEval-Cpp} task, the mean \discore{} for C++ training data is -2.62$e$-04, which is among the lowest.
However, we observed a crucial difference in the variance: the standard deviation of \discore{} for same-language data is notably higher than that of data from other languages. As shown in Table~\ref{tab:language}, the standard deviation for Python data is significantly greater in both \textit{HumanEval-Python} and \textit{DS-1000} tasks. Similarly, SQL data shows the highest standard deviation in the \textit{Bird-SQL} task. This results in a "short and fat" distribution for the influence scores of same-language training data, meaning these distributions are flatter and wider.

These findings suggest that simply increasing the volume of same-language training data is an inefficient strategy. Instead, the higher standard deviation indicates that same-language data contains a wider range of influence scores, including both highly beneficial and highly detrimental samples. Therefore, effective filtering to identify and select high-quality same-language data is crucial for optimizing model performance on language-specific tasks.

\mybox{Finding 5: While same-language data offers no significant average benefit over other languages, it exhibits a much wider variance in \discore{}. This indicates the presence of both highly beneficial and highly detrimental samples, making targeted filtering essential for improving performance on language-specific tasks.}

\begin{figure}
  \centering
  \begin{subfigure}[b]{1\linewidth}
  \centering
  \includegraphics[width=1\linewidth]{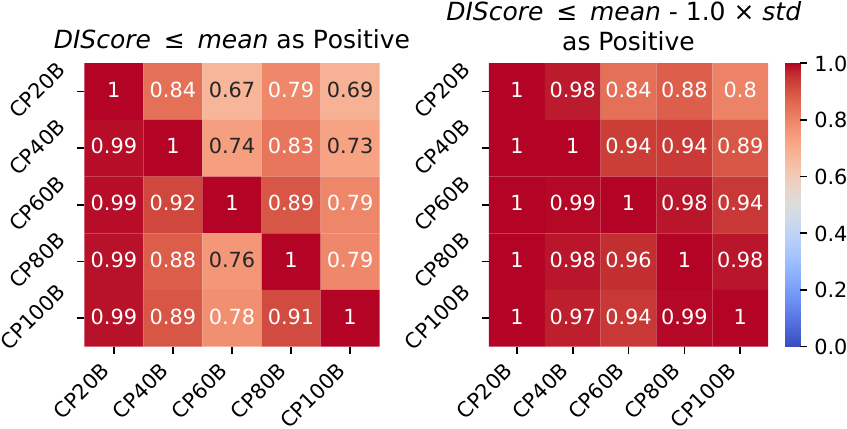}
  \caption{Precision negative sample between different checkpoints.}
  \label{fig:corr1}
  \end{subfigure}
  \begin{subfigure}[b]{1\linewidth}
  \centering
  \includegraphics[width=1\linewidth]{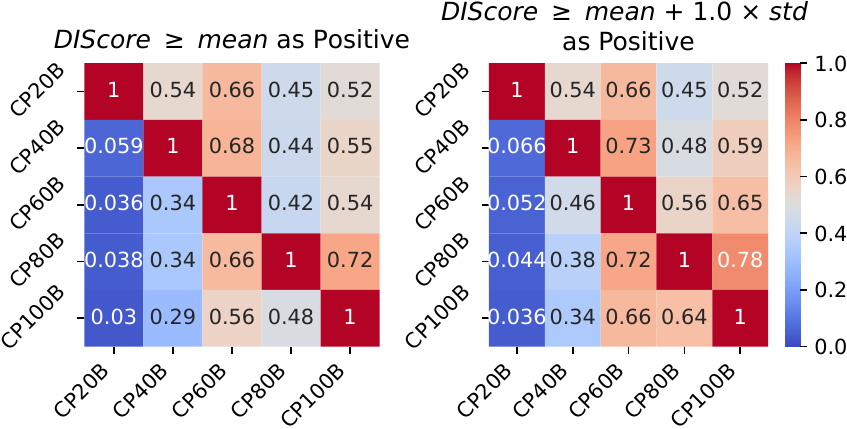}
  \caption{Precision positive sample between different checkpoints.}
  \label{fig:corr2}
  \end{subfigure}
    \caption{Precision of negative and positive samples across \codellm{} checkpoints.}
    \label{fig:corr}
\end{figure}

As previously noted, the \discore{} of most training data is concentrated around zero, meaning that for a large portion of samples, their \discore{} is not particularly significant. In other words, for samples with \discore{} near 0, there is little distinction between beneficial and harmful data, and the effect of such data appears somewhat random.
Given this, we shift our focus to the samples at the extremes of the distribution (i.e., those that deviate significantly from the $mean$). Specifically, we analyze samples beyond the $mean$ (i.e., mean of \discore{}), $mean$$\pm$$0.5$$\times$$std$ (i.e., standard-deviations of \discore{}), and $mean$$\pm$$1.0$$\times$$std$. Instead of using Spearman correlation analysis, which measures linear relationships, we concentrate on whether data samples consistently benefit the training process across different model checkpoints. To do this, we compute the positive sample precision and negative sample precision between various training stages, quantifying how consistently a sample is judged as beneficial or harmful across checkpoints.

\begin{sloppypar}
Figure~\ref{fig:corr} illustrates the positive and negative sample precision between 5 different model checkpoints, under different $standard$-$deviations$ ranges of \discore{}. For example, consider the first row and last column of the middle sub-figure in Figure~\ref{fig:corr1}, which displays the negative sample precision of \codellm{}\codefont{-CP20B} relative to \codellm{}\codefont{-CP100B}. The data reveals that 79\% of the samples identified as negative (i.e., \discore{} $\leq$ $mean$-$0.5$$\times$$std$) in \codellm{}\codefont{-CP20B} are also identified as negative in \codellm{}\codefont{-CP100B}. The corresponding positive sample precision can be seen in Figure~\ref{fig:corr2}.
We observe that as we consider samples with more extreme \discore{} (i.e., those further from the mean of \discore{}), the precision of both positive and negative samples across checkpoints improves significantly. For instance, between \codellm{}\codefont{-CP20B} and \codellm{}\codefont{-CP100B}, when considering samples with \discore{} less than the $mean$ as negative influence (i.e., first sub-figure in Figure~\ref{fig:corr1}), the negative sample precision is 0.69. However, when we narrow the selection to samples less than $mean$-$0.5$$\times$$std$ as negative (i.e., second sub-figure in Figure~\ref{fig:corr1}), the negative sample precision increases to 0.88. A similar pattern is seen in positive samples precision. For example, the positive sample precision between \codellm{}\codefont{-CP40B} and \codellm{}\codefont{-CP100B} increases from 0.68 to 0.73 in Figure~\ref{fig:corr2}. However, this trend is more pronounced for negative samples than for positive ones.
\end{sloppypar}

This observation suggests that negative samples tend to exhibit more consistency across different training stages compared to positive samples. In other words, a data sample identified as negative at one checkpoint is more likely to be consistently identified as negative at other checkpoints, especially when focusing on samples with more extreme negative \discore{}. This trend could inform the development of more effective data filtering strategies, particularly for identifying and removing harmful data.

\mybox{Finding 6: Samples with more extreme \discore{} show more consistent trends across different training stages, while intermediate samples exhibit more randomness. These intermediate samples may be beneficial at one stage but not necessarily at others.}

\subsubsection{RQ3: How do mainstream data filtering methods compare with the data-influence-score filtering method?\\}

\begin{figure*}
  \centering
  \includegraphics[width=\linewidth]{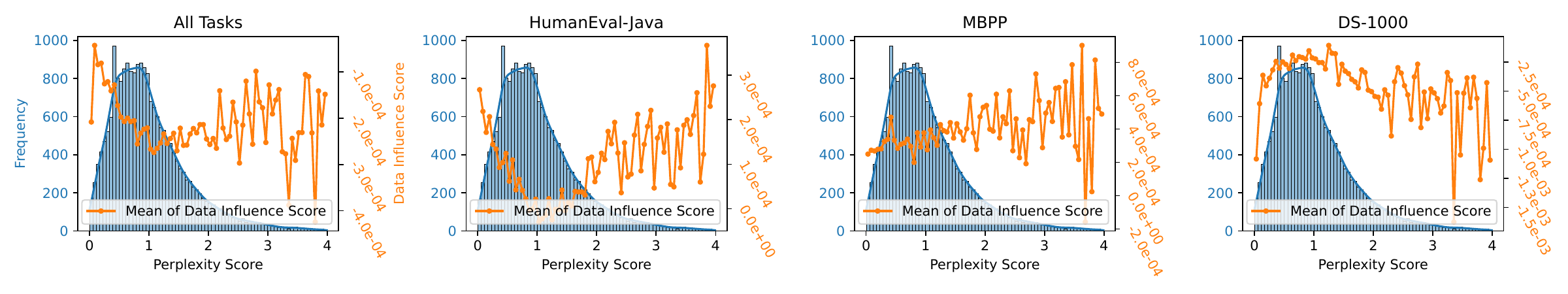}
  \caption{Distribution of perplexity values and their corresponding average \discore{} across different tasks.}
  \label{fig:score1}
\end{figure*}

\begin{figure}
\centering
\includegraphics[width=1\linewidth]{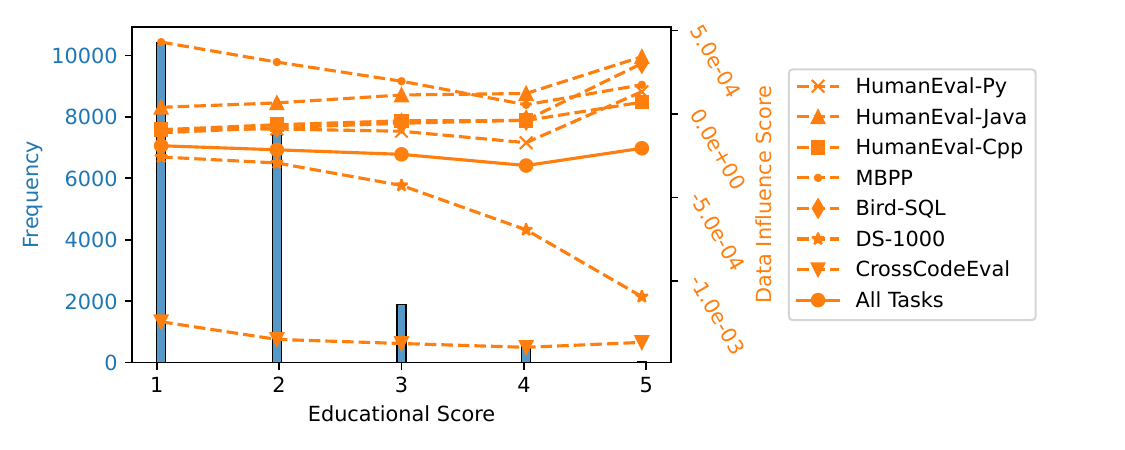}
\caption{Distribution of educational scores and their corresponding average \discore{}.}
\label{fig:score2}
\end{figure}

To better understand the characteristics of \discore{}, we analyzed the samples selected by two commonly used data filtering strategies, i.e., perplexity-based filtering and LLM scoring.

\ptextbf{Perplexity-Based Filtering}

Firstly, for the perplexity-based filtering, we followed the methodology from prior work by using CodeLLM-CP100B as the scoring model. We computed the perplexity for each training sample from Section~\ref{sec:subject2} and plotted histograms of the perplexity values. Each histogram bin was treated as a cluster, and we calculated the average \discore{} for the samples within each bin. Figure~\ref{fig:score1} illustrates these perplexity histograms along with their corresponding average \discore{} across different tasks.
From Figure~\ref{fig:score1}, it is evident that the majority of data samples have perplexity values concentrated between 0 and 4. Interestingly, for all the tasks, the \discore{} generally exhibits a non-linear pattern, that decreases initially and then gradually increases as perplexity values rise. This trend is particularly pronounced in the \textit{HumanEval-Java} task, suggesting that samples with both low and high perplexity may be more beneficial for training, while those with medium perplexity tend to be less impactful.
However, this pattern is not consistent across all tasks. For example, in the \textit{DS-1000} task, \discore{} first increases and then slowly declines as perplexity rises, whereas in the \textit{MBPP} task, the \discore{} shows a steady upward trend. A closer look at the absolute values of \discore{} within each perplexity interval reveals that most values range between -$4e$-$4$ and 0, indicating no substantial variation in \discore{} across samples with different perplexity levels. Therefore, perplexity-based methods appear to be ineffective in identifying samples with higher \discore{}.

\ptextbf{LLM-Based Scoring}

Next, we investigated the \discore{} characteristics of samples selected by an LLM-based scoring approach. As in previous work~\cite{phi1}, we used GPT-4o~\cite{chatgpt} to score the educational value of each sample, where the model assigns a score between 1 and 5. A higher score reflects the model's assessment of the sample's potential benefit for training. Figure~\ref{fig:score2} presents the distribution of educational scores for the training data, along with their corresponding average \discore{} for each task.
Interestingly, the majority of samples received educational scores of 1 or 2, accounting for up to 90\% of all data points. Due to the limited number of samples with a score of 5, we omit detailed discussion of this category. Surprisingly, the \discore{} tends to decrease as the educational score increases from 1 to 4. This suggests that samples deemed more beneficial for training by the LLM do not necessarily correspond to those with higher \discore{}.
Moreover, the trends vary across different tasks. For instance, in the \textit{HumanEval-Java} task, \discore{} increases as the educational score rises. However, a more detailed analysis of the absolute values of \discore{} for each educational score reveals that most values fall between -$3e$-$4$ and -$2e$-$4$, which is not significantly different in \discore{} across samples with different educational score. Consequently, LLM-based scoring methods also struggle to effectively differentiate between samples with high and low \discore{}.

\mybox{Finding 7: While perplexity-based and LLM-based scoring methods exhibit different behaviors across tasks, neither method effectively distinguishes between samples with high and low \discore{}.}

\subsubsection{RQ4: How effective is the prediction-based data-influence-score filtering method?\\}

Calculating the \discore{} for every sample in a large dataset requires performing one-step training for each sample and then evaluating the model on the validation set. This process is computationally expensive and practically infeasible for large datasets. To mitigate the high cost of obtaining \discore{}, existing methods (such as MATES~\cite{mates}) typically rely on training a smaller model (i.e., RoBERTa~\cite{roberta}) to predict the \discore{} of each sample.

In this study, we adopt a similar approach to evaluate the effectiveness of prediction-based \discore{} filtering method in code generation tasks. Specifically, after training \codellm{} on a substantial amount of data (e.g., 20B code tokens), we perform one-step training on a smaller subset (around 20,000 samples) to obtain oracle \discore{} labels for these samples. Next, we train a RoBERTa-Base model~\cite{roberta} using the labeled subset to predict the \discore{} for the entire training dataset. Based on these predicted scores, we then select a new batch of data (another 20 billion tokens) from the remaining unprocessed training data for \codellm{} to continue training. This process is iterative. After completing training on the newly selected batch, we repeat the following steps: (1) Use the updated \codellm{} to label \discore{} values for a new small subset of data. (2) Train the RoBERTa-Base model on this labeled data with a regression objective. (3) Use the trained RoBERTa-Base model to predict \discore{} values for the entire training dataset. (4) Select another batch of training data based on the predicted scores. (5) Continue training \codellm{} on the newly selected data. This cycle is repeated until the entire training process is complete.

To ensure comparability with previous studies~\cite{mates}, we followed the same hyperparameter settings. The training process involved a total of 100B code tokens, with 20B tokens selected in each iteration. In each cycle, we labeled 20,000 samples with oracle \discore{}. We used the validation set mentioned in Section~\ref{sec:subject2} to evaluate \discore{} and assessed the model's performance on the downstream tasks described in Section~\ref{sec:subject2}. Since the validation set shares the same distribution as the evaluation data, this setup makes our experiment align with an in-distribution \discore{} filtering scheme.

\begin{figure}
  \centering
  \includegraphics[width=0.9\linewidth]{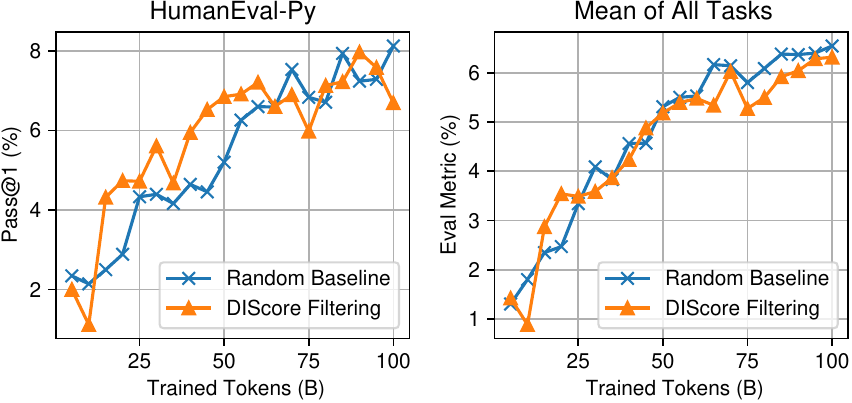}
  \caption{Performance comparison of \discore{} filtering versus random selection under programming tasks.}
  \label{fig:train}
\end{figure}

\begin{figure}
  \centering
  \begin{subfigure}[b]{0.49\linewidth}
  \centering
  \includegraphics[width=1\linewidth]{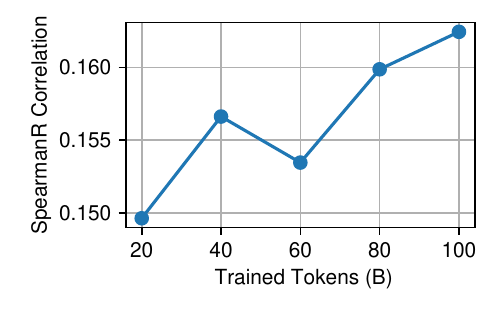}
  \caption{Spearman rank correlation at different training stages.}
  \label{fig:pred1}
  \end{subfigure}
  \begin{subfigure}[b]{0.49\linewidth}
  \centering
  \includegraphics[width=1\linewidth]{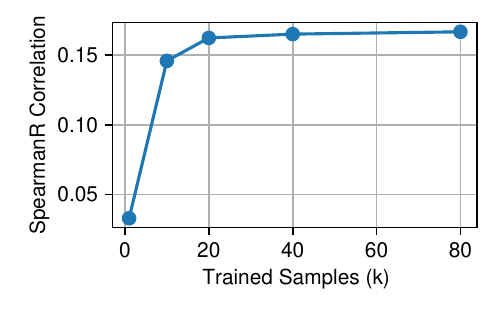}
  \caption{Spearman rank correlation with varying numbers of oracle \discore{} samples.}
  \label{fig:pred2}
  \end{subfigure}
    \caption{Analysis of prediction accuracy for \discore{}: (a) correlation at different training stages; (b) correlation with varying numbers of training samples.}
    \label{fig:pred}
\end{figure}

Figure~\ref{fig:train} presents the training results, indicate that the \discore{} filtering method based on small model predictions does not significantly outperform random selection across various programming tasks. This suggests that the \discore{} predicted by the small model does not meaningfully improve the model's performance on downstream tasks.
To gain a deeper insight into this phenomenon, we further analyzed the accuracy of the small model's predictions at different stages of training. In Figure~\ref{fig:pred1}, we present the Spearman rank correlation coefficients between the small model's predictions and the oracle \discore{} at various stages of training. The results reveal that the small model struggles to accurately predict oracle \discore{}, with the highest correlation reaching only 0.1624, that far from a significant level. This indicates that the small model has limited capability to differentiate between high and low \discore{} data.
Additionally, in Figure~\ref{fig:pred2}, we examine how the Spearman rank correlation coefficient changes at \codellm{}\codefont{-CP100B} when using different numbers of oracle \discore{} samples for training. Even when training with 80,000 samples, the correlation coefficient only reaches 0.1634, demonstrating that increasing the number of labeled samples does not significantly enhance the small model's predictive accuracy.
We hypothesize that this poor performance stems from several factors. First, code generation tasks are highly complex and specialized, with \discore{} being shaped by numerous factors that are difficult for a small model to capture. Second, as noted in Finding 4, the \discore{} of the same training data can vary widely across different programming tasks, further complicating prediction and limiting the small model's ability to generalize. Finally, small models have limited representational and comprehension capacities, making it challenging for them to accurately model the intricate relationships underlying \discore{}.

Based on the experimental results and analysis above, we conclude that using small models to predict \discore{} in code generation tasks is ineffective. Due to the low prediction accuracy of small models, the quality of training data selected based on their predictions is not significantly better than that of randomly selected data. This suggests that, for complex tasks like code generation, relying on small models to approximate \discore{} may not be a viable strategy, and more effective alternatives should be explored.

\mybox{Finding 8: Even in an in-distribution setting, \discore{} filtering strategy based on small model predictions are ineffective. This is primarily due to the small models' low accuracy in predicting data influence, which makes it difficult to select beneficial training data.}

\section{Implications and Discussions}

Our study offers several key insights and practical guidelines for future research on pre-training data selection for Code-LLMs.

\ptextbf{A More Accurate \textbf{\textit{DIScore}} Computation Method}
Finding 1 confirms that computing \discore{} with \vsloss{} generally aligns with downstream task metrics, and filtering based on it improves model performance. However, a gap remains between this measurement and actual downstream impact, which suggests that \vsloss{} alone may not fully capture a data point's value. Consequently, future work should explore more accurate methods for calculating \discore{} to better assess the true influence of training data.

\ptextbf{Improving \textbf{\textit{DIScore}} Prediction Methods:}
As highlighted in Finding 8, current \discore{} prediction methods that rely on small proxy models~\cite{mates,dsdm} suffer from low accuracy. This indicates that simple proxy models are insufficient for capturing the complex influence patterns in code data. Consequently, this prediction bottleneck currently limits the scalability of our method to massive datasets. Future research needs to investigate more sophisticated prediction strategies, such as using larger or more specialized proxy models, or incorporating additional data features to enhance selection accuracy.

\ptextbf{Re-evaluate the Effectiveness of Traditional Data Selection Methods in Code Data:}
According to Finding 7, traditional selection methods like perplexity~\cite{ppl} and LLM scoring~\cite{askllm,qurating,phi1} are ineffective at identifying high-\discore{} samples in code data. This indicates that strategies common in NLP may not be suitable for code, as they fail to assess a sample's true contribution. This underscores the need to re-evaluate these methods and develop a code-specific evaluation system that considers attributes like correctness, executability, complexity, and task relevance.

\ptextbf{Investigating Task-Specific \textbf{\textit{DIScore}} to Build a Generalized Validation Set:}
Finding 4 reveals that the influence of training data is highly task-specific. While constructing task-specific validation sets can boost performance in targeted domains, it introduces the risk of "overfitting" to benchmarks or reducing data diversity. Therefore, constructing a generalized validation set that balances specific task requirements with broad reasoning capabilities is crucial to prevent the model from becoming overly specialized or "leaking" benchmark patterns and would improve the model's overall generalization and performance.

\ptextbf{Focus on the Consistency of Extreme Samples:}
Finding 6 indicates that samples with extreme \discore{} (i.e., highly positive or negative) exhibit consistent influence throughout training, whereas moderate-influence samples are more variable. This suggests that data optimization should prioritize these consistently impactful samples. Moreover, identifying and removing consistently detrimental samples can prevent the model from learning harmful patterns. Thus, focusing on the consistency of extreme samples is a crucial strategy for building high-quality training datasets.

\ptextbf{Fine-Tuning with High-Value Data:}
While applying \textbf{\discore{}} filtering to massive pre-training corpora is computationally intensive, our findings suggest immediate practical value in selecting high-quality data for ``Instruction Tuning'' (SFT) or identifying ``Seed Data'' for continued pre-training. In these scenarios, the dataset size is manageable, making the high cost of \discore{} calculation justifiable. By acting as a ``Gold Standard'' filter, our method can curate high-quality subsets that maximize efficiency in fine-tuning stages.

\ptextbf{Limitations on Computation Overhead and Execution Time:}
We acknowledge that a significant limitation of our current approach is the computational overhead. Calculating \discore{} based on \vsloss{} requires performing a forward and backward pass for each sample against the validation set. This execution time scales linearly with the size of the validation set and the number of training samples, making it costly for trillion-token scale pre-training without optimization. Future work should focus on reducing this runtime overhead, potentially through influence approximation techniques or more efficient gradient analysis.

\section{Threats to Validity}

\parabf{Threats to Internal Validity.}  
The threats to internal validity mainly lie in the potential bugs in our implementation. To mitigate these risks, the authors have meticulously reviewed the code and scripts. Furthermore, we released the code and scripts in ~\cite{codescripts} for public scrutiny and also facilitate independent verification of our findings. 


\parabf{Threats to External Validity.}  
These threats mainly lie in two aspects. First, regarding the data filtering methods, we have conducted an extensive literature review and believe that the filtering methods used are representative. Second, regarding the model scale, our empirical study was conducted on a 1B parameter model due to computational resource constraints.

\parabf{Threats to Construct Validity.}
This threats primarily arise from the downstream tasks used in our evaluations. To mitigate these threats, we have employed a range of widely-recognized programming tasks (e.g., HumanEval, MBPP, CrossCodeEval) to assess the practical performance of the real-world tasks.

\section{Conclusion}
In this paper, we explored the application of \discore{} filtering for optimizing pre-training data in Code-LLMs. By introducing a novel method for calculating \discore{} based on \vsloss{} for generative programming tasks, we demonstrated that this approach can significantly enhance model performance across various programming tasks.
Our extensive empirical study, using a 1B-parameter CodeLLM pre-trained on 100 billion code tokens, revealed key insights into how the characteristics of beneficial training data evolve over different training stages and vary across programming tasks. We also found that predicting the oracle \discore{} accurately remains challenging, particularly when using smaller models for approximation.
Overall, our findings underscore the importance of tailored data filtering strategies for code-specific datasets and provide a solid foundation for future research aimed at optimizing pre-training data for Code-LLMs.


\bibliographystyle{ACM-Reference-Format}
\bibliography{ref}

@article{codex,
  author       = {Mark Chen and
                  Jerry Tworek and
                  Heewoo Jun and
                  Qiming Yuan and
                  Henrique Pond{\'{e}} de Oliveira Pinto and
                  Jared Kaplan and
                  Harrison Edwards and
                  Yuri Burda and
                  Nicholas Joseph and
                  Greg Brockman and
                  Alex Ray and
                  Raul Puri and
                  Gretchen Krueger and
                  Michael Petrov and
                  Heidy Khlaaf and
                  Girish Sastry and
                  Pamela Mishkin and
                  Brooke Chan and
                  Scott Gray and
                  Nick Ryder and
                  Mikhail Pavlov and
                  Alethea Power and
                  Lukasz Kaiser and
                  Mohammad Bavarian and
                  Clemens Winter and
                  Philippe Tillet and
                  Felipe Petroski Such and
                  Dave Cummings and
                  Matthias Plappert and
                  Fotios Chantzis and
                  Elizabeth Barnes and
                  Ariel Herbert{-}Voss and
                  William Hebgen Guss and
                  Alex Nichol and
                  Alex Paino and
                  Nikolas Tezak and
                  Jie Tang and
                  Igor Babuschkin and
                  Suchir Balaji and
                  Shantanu Jain and
                  William Saunders and
                  Christopher Hesse and
                  Andrew N. Carr and
                  Jan Leike and
                  Joshua Achiam and
                  Vedant Misra and
                  Evan Morikawa and
                  Alec Radford and
                  Matthew Knight and
                  Miles Brundage and
                  Mira Murati and
                  Katie Mayer and
                  Peter Welinder and
                  Bob McGrew and
                  Dario Amodei and
                  Sam McCandlish and
                  Ilya Sutskever and
                  Wojciech Zaremba},
  title        = {Evaluating Large Language Models Trained on Code},
  journal      = {CoRR},
  volume       = {abs/2107.03374},
  year         = {2021},
  url          = {https://arxiv.org/abs/2107.03374},
  eprinttype    = {arXiv},
  eprint       = {2107.03374},
  bibsource    = {dblp computer science bibliography, https://dblp.org}
}

@article{codereval,
  author       = {Hao Yu and
                  Bo Shen and
                  Dezhi Ran and
                  Jiaxin Zhang and
                  Qi Zhang and
                  Yuchi Ma and
                  Guangtai Liang and
                  Ying Li and
                  Tao Xie and
                  Qianxiang Wang},
  title        = {CoderEval: {A} Benchmark of Pragmatic Code Generation with Generative
                  Pre-trained Models},
  journal      = {CoRR},
  volume       = {abs/2302.00288},
  year         = {2023},
  url          = {https://doi.org/10.48550/arXiv.2302.00288},
  doi          = {10.48550/ARXIV.2302.00288},
  eprinttype    = {arXiv},
  eprint       = {2302.00288},
  bibsource    = {dblp computer science bibliography, https://dblp.org}
}

@misc{chatgpt,
    title = "ChatGPT",
    howpublished = "Website",
    year = {2023},
    note = {\url{https://openai.com/blog/chatgpt}}
}

@article{starcoder,
  author       = {Raymond Li and
                  Loubna Ben Allal and
                  Yangtian Zi and
                  Niklas Muennighoff and
                  Denis Kocetkov and
                  Chenghao Mou and
                  Marc Marone and
                  Christopher Akiki and
                  Jia Li and
                  Jenny Chim and
                  Qian Liu and
                  Evgenii Zheltonozhskii and
                  Terry Yue Zhuo and
                  Thomas Wang and
                  Olivier Dehaene and
                  Mishig Davaadorj and
                  Joel Lamy{-}Poirier and
                  Jo{\~{a}}o Monteiro and
                  Oleh Shliazhko and
                  Nicolas Gontier and
                  Nicholas Meade and
                  Armel Zebaze and
                  Ming{-}Ho Yee and
                  Logesh Kumar Umapathi and
                  Jian Zhu and
                  Benjamin Lipkin and
                  Muhtasham Oblokulov and
                  Zhiruo Wang and
                  Rudra Murthy V and
                  Jason Stillerman and
                  Siva Sankalp Patel and
                  Dmitry Abulkhanov and
                  Marco Zocca and
                  Manan Dey and
                  Zhihan Zhang and
                  Nour Moustafa{-}Fahmy and
                  Urvashi Bhattacharyya and
                  Wenhao Yu and
                  Swayam Singh and
                  Sasha Luccioni and
                  Paulo Villegas and
                  Maxim Kunakov and
                  Fedor Zhdanov and
                  Manuel Romero and
                  Tony Lee and
                  Nadav Timor and
                  Jennifer Ding and
                  Claire Schlesinger and
                  Hailey Schoelkopf and
                  Jan Ebert and
                  Tri Dao and
                  Mayank Mishra and
                  Alex Gu and
                  Jennifer Robinson and
                  Carolyn Jane Anderson and
                  Brendan Dolan{-}Gavitt and
                  Danish Contractor and
                  Siva Reddy and
                  Daniel Fried and
                  Dzmitry Bahdanau and
                  Yacine Jernite and
                  Carlos Mu{\~{n}}oz Ferrandis and
                  Sean Hughes and
                  Thomas Wolf and
                  Arjun Guha and
                  Leandro von Werra and
                  Harm de Vries},
  title        = {StarCoder: may the source be with you!},
  journal      = {CoRR},
  volume       = {abs/2305.06161},
  year         = {2023},
  url          = {https://doi.org/10.48550/arXiv.2305.06161},
  doi          = {10.48550/ARXIV.2305.06161},
  eprinttype    = {arXiv},
  eprint       = {2305.06161},
  bibsource    = {dblp computer science bibliography, https://dblp.org}
}

@article{codellama,
  author       = {Baptiste Rozi{\`{e}}re and
                  Jonas Gehring and
                  Fabian Gloeckle and
                  Sten Sootla and
                  Itai Gat and
                  Xiaoqing Ellen Tan and
                  Yossi Adi and
                  Jingyu Liu and
                  Tal Remez and
                  J{\'{e}}r{\'{e}}my Rapin and
                  Artyom Kozhevnikov and
                  Ivan Evtimov and
                  Joanna Bitton and
                  Manish Bhatt and
                  Cristian Canton{-}Ferrer and
                  Aaron Grattafiori and
                  Wenhan Xiong and
                  Alexandre D{\'{e}}fossez and
                  Jade Copet and
                  Faisal Azhar and
                  Hugo Touvron and
                  Louis Martin and
                  Nicolas Usunier and
                  Thomas Scialom and
                  Gabriel Synnaeve},
  title        = {Code Llama: Open Foundation Models for Code},
  journal      = {CoRR},
  volume       = {abs/2308.12950},
  year         = {2023},
  url          = {https://doi.org/10.48550/arXiv.2308.12950},
  doi          = {10.48550/ARXIV.2308.12950},
  eprinttype    = {arXiv},
  eprint       = {2308.12950},
  bibsource    = {dblp computer science bibliography, https://dblp.org}
}

@article{nl2fix,
  author       = {Sarah Fakhoury and
                  Saikat Chakraborty and
                  Madan Musuvathi and
                  Shuvendu K. Lahiri},
  title        = {Towards Generating Functionally Correct Code Edits from Natural Language
                  Issue Descriptions},
  journal      = {CoRR},
  volume       = {abs/2304.03816},
  year         = {2023},
  url          = {https://doi.org/10.48550/arXiv.2304.03816},
  doi          = {10.48550/ARXIV.2304.03816},
  eprinttype    = {arXiv},
  eprint       = {2304.03816},
  bibsource    = {dblp computer science bibliography, https://dblp.org}
}

@inproceedings{transformer,
  author       = {Ashish Vaswani and
                  Noam Shazeer and
                  Niki Parmar and
                  Jakob Uszkoreit and
                  Llion Jones and
                  Aidan N. Gomez and
                  Lukasz Kaiser and
                  Illia Polosukhin},
  editor       = {Isabelle Guyon and
                  Ulrike von Luxburg and
                  Samy Bengio and
                  Hanna M. Wallach and
                  Rob Fergus and
                  S. V. N. Vishwanathan and
                  Roman Garnett},
  title        = {Attention is All you Need},
  booktitle    = {Advances in Neural Information Processing Systems 30: Annual Conference
                  on Neural Information Processing Systems 2017, December 4-9, 2017,
                  Long Beach, CA, {USA}},
  pages        = {5998--6008},
  year         = {2017},
  url          = {https://proceedings.neurips.cc/paper/2017/hash/3f5ee243547dee91fbd053c1c4a845aa-Abstract.html},
  bibsource    = {dblp computer science bibliography, https://dblp.org}
}

@inproceedings{coderujb,
  author       = {Zhengran Zeng and
                  Yidong Wang and
                  Rui Xie and
                  Wei Ye and
                  Shikun Zhang},
  editor       = {Maria Christakis and
                  Michael Pradel},
  title        = {CoderUJB: An Executable and Unified Java Benchmark for Practical Programming
                  Scenarios},
  booktitle    = {Proceedings of the 33rd {ACM} {SIGSOFT} International Symposium on
                  Software Testing and Analysis, {ISSTA} 2024, Vienna, Austria, September
                  16-20, 2024},
  pages        = {124--136},
  publisher    = {{ACM}},
  year         = {2024},
  url          = {https://doi.org/10.1145/3650212.3652115},
  doi          = {10.1145/3650212.3652115},
  bibsource    = {dblp computer science bibliography, https://dblp.org}
}

@inproceedings{swebench,
  author       = {Carlos E. Jimenez and
                  John Yang and
                  Alexander Wettig and
                  Shunyu Yao and
                  Kexin Pei and
                  Ofir Press and
                  Karthik R. Narasimhan},
  title        = {SWE-bench: Can Language Models Resolve Real-world Github Issues?},
  booktitle    = {The Twelfth International Conference on Learning Representations,
                  {ICLR} 2024, Vienna, Austria, May 7-11, 2024},
  publisher    = {OpenReview.net},
  year         = {2024},
  url          = {https://openreview.net/forum?id=VTF8yNQM66},
  bibsource    = {dblp computer science bibliography, https://dblp.org}
}

@article{codeagentsurvey,
  author       = {Junwei Liu and
                  Kaixin Wang and
                  Yixuan Chen and
                  Xin Peng and
                  Zhenpeng Chen and
                  Lingming Zhang and
                  Yiling Lou},
  title        = {Large Language Model-Based Agents for Software Engineering: {A} Survey},
  journal      = {CoRR},
  volume       = {abs/2409.02977},
  year         = {2024},
  url          = {https://doi.org/10.48550/arXiv.2409.02977},
  doi          = {10.48550/ARXIV.2409.02977},
  eprinttype    = {arXiv},
  eprint       = {2409.02977},
  bibsource    = {dblp computer science bibliography, https://dblp.org}
}

@inproceedings{codesurvey,
  author       = {Angela Fan and
                  Beliz Gokkaya and
                  Mark Harman and
                  Mitya Lyubarskiy and
                  Shubho Sengupta and
                  Shin Yoo and
                  Jie M. Zhang},
  title        = {Large Language Models for Software Engineering: Survey and Open Problems},
  booktitle    = {{IEEE/ACM} International Conference on Software Engineering: Future
                  of Software Engineering, ICSE-FoSE 2023, Melbourne, Australia, May
                  14-20, 2023},
  pages        = {31--53},
  publisher    = {{IEEE}},
  year         = {2023},
  url          = {https://doi.org/10.1109/ICSE-FoSE59343.2023.00008},
  doi          = {10.1109/ICSE-FOSE59343.2023.00008},
  bibsource    = {dblp computer science bibliography, https://dblp.org}
}

@article{starcoder2,
  author       = {Anton Lozhkov and
                  Raymond Li and
                  Loubna Ben Allal and
                  Federico Cassano and
                  Joel Lamy{-}Poirier and
                  Nouamane Tazi and
                  Ao Tang and
                  Dmytro Pykhtar and
                  Jiawei Liu and
                  Yuxiang Wei and
                  Tianyang Liu and
                  Max Tian and
                  Denis Kocetkov and
                  Arthur Zucker and
                  Younes Belkada and
                  Zijian Wang and
                  Qian Liu and
                  Dmitry Abulkhanov and
                  Indraneil Paul and
                  Zhuang Li and
                  Wen{-}Ding Li and
                  Megan Risdal and
                  Jia Li and
                  Jian Zhu and
                  Terry Yue Zhuo and
                  Evgenii Zheltonozhskii and
                  Nii Osae Osae Dade and
                  Wenhao Yu and
                  Lucas Krau{\ss} and
                  Naman Jain and
                  Yixuan Su and
                  Xuanli He and
                  Manan Dey and
                  Edoardo Abati and
                  Yekun Chai and
                  Niklas Muennighoff and
                  Xiangru Tang and
                  Muhtasham Oblokulov and
                  Christopher Akiki and
                  Marc Marone and
                  Chenghao Mou and
                  Mayank Mishra and
                  Alex Gu and
                  Binyuan Hui and
                  Tri Dao and
                  Armel Zebaze and
                  Olivier Dehaene and
                  Nicolas Patry and
                  Canwen Xu and
                  Julian J. McAuley and
                  Han Hu and
                  Torsten Scholak and
                  S{\'{e}}bastien Paquet and
                  Jennifer Robinson and
                  Carolyn Jane Anderson and
                  Nicolas Chapados and
                  et al.},
  title        = {StarCoder 2 and The Stack v2: The Next Generation},
  journal      = {CoRR},
  volume       = {abs/2402.19173},
  year         = {2024},
  url          = {https://doi.org/10.48550/arXiv.2402.19173},
  doi          = {10.48550/ARXIV.2402.19173},
  eprinttype    = {arXiv},
  eprint       = {2402.19173},
  bibsource    = {dblp computer science bibliography, https://dblp.org}
}

@article{llmsurvey,
  title={A survey of large language models},
  author={Zhao, Wayne Xin and Zhou, Kun and Li, Junyi and Tang, Tianyi and Wang, Xiaolei and Hou, Yupeng and Min, Yingqian and Zhang, Beichen and Zhang, Junjie and Dong, Zican and others},
  journal={arXiv preprint arXiv:2303.18223},
  year={2023}
}

@article{codellmsurvey,
  author       = {Ziyin Zhang and
                  Chaoyu Chen and
                  Bingchang Liu and
                  Cong Liao and
                  Zi Gong and
                  Hang Yu and
                  Jianguo Li and
                  Rui Wang},
  title        = {A Survey on Language Models for Code},
  journal      = {CoRR},
  volume       = {abs/2311.07989},
  year         = {2023},
  url          = {https://doi.org/10.48550/arXiv.2311.07989},
  doi          = {10.48550/ARXIV.2311.07989},
  eprinttype    = {arXiv},
  eprint       = {2311.07989},
  bibsource    = {dblp computer science bibliography, https://dblp.org}
}

@article{phi1,
  title={Textbooks are all you need},
  author={Gunasekar, Suriya and Zhang, Yi and Aneja, Jyoti and Mendes, Caio C{\'e}sar Teodoro and Del Giorno, Allie and Gopi, Sivakanth and Javaheripi, Mojan and Kauffmann, Piero and de Rosa, Gustavo and Saarikivi, Olli and others},
  journal={arXiv preprint arXiv:2306.11644},
  year={2023}
}

@article{santacoder,
  title={SantaCoder: don't reach for the stars!},
  author={Allal, Loubna Ben and Li, Raymond and Kocetkov, Denis and Mou, Chenghao and Akiki, Christopher and Ferrandis, Carlos Munoz and Muennighoff, Niklas and Mishra, Mayank and Gu, Alex and Dey, Manan and others},
  journal={arXiv preprint arXiv:2301.03988},
  year={2023}
}

@article{ppl,
  title={Perplexed by Perplexity: Perplexity-Based Data Pruning With Small Reference Models},
  author={Ankner, Zachary and Blakeney, Cody and Sreenivasan, Kartik and Marion, Max and Leavitt, Matthew L and Paul, Mansheej},
  journal={arXiv preprint arXiv:2405.20541},
  year={2024}
}

@article{qurating,
  title={Qurating: Selecting high-quality data for training language models},
  author={Wettig, Alexander and Gupta, Aatmik and Malik, Saumya and Chen, Danqi},
  journal={arXiv preprint arXiv:2402.09739},
  year={2024}
}

@article{askllm,
  title={How to Train Data-Efficient LLMs},
  author={Sachdeva, Noveen and Coleman, Benjamin and Kang, Wang-Cheng and Ni, Jianmo and Hong, Lichan and Chi, Ed H and Caverlee, James and McAuley, Julian and Cheng, Derek Zhiyuan},
  journal={arXiv preprint arXiv:2402.09668},
  year={2024}
}

@article{mates,
  title={MATES: Model-Aware Data Selection for Efficient Pretraining with Data Influence Models},
  author={Yu, Zichun and Das, Spandan and Xiong, Chenyan},
  journal={arXiv preprint arXiv:2406.06046},
  year={2024}
}

@article{minicpm,
  title={Minicpm: Unveiling the potential of small language models with scalable training strategies},
  author={Hu, Shengding and Tu, Yuge and Han, Xu and He, Chaoqun and Cui, Ganqu and Long, Xiang and Zheng, Zhi and Fang, Yewei and Huang, Yuxiang and Zhao, Weilin and others},
  journal={arXiv preprint arXiv:2404.06395},
  year={2024}
}

@article{dsdm,
  title={Dsdm: Model-aware dataset selection with datamodels},
  author={Engstrom, Logan and Feldmann, Axel and Madry, Aleksander},
  journal={arXiv preprint arXiv:2401.12926},
  year={2024}
}

@article{deepseekcoder,
  title={DeepSeek-Coder: When the Large Language Model Meets Programming--The Rise of Code Intelligence},
  author={Guo, Daya and Zhu, Qihao and Yang, Dejian and Xie, Zhenda and Dong, Kai and Zhang, Wentao and Chen, Guanting and Bi, Xiao and Wu, Yu and Li, YK and others},
  journal={arXiv preprint arXiv:2401.14196},
  year={2024}
}

@article{qwen2,
  title={Qwen2. 5-coder technical report},
  author={Hui, Binyuan and Yang, Jian and Cui, Zeyu and Yang, Jiaxi and Liu, Dayiheng and Zhang, Lei and Liu, Tianyu and Zhang, Jiajun and Yu, Bowen and Dang, Kai and others},
  journal={arXiv preprint arXiv:2409.12186},
  year={2024}
}

@article{ppl0,
  title={Exploring the limits of transfer learning with a unified text-to-text transformer},
  author={Raffel, Colin and Shazeer, Noam and Roberts, Adam and Lee, Katherine and Narang, Sharan and Matena, Michael and Zhou, Yanqi and Li, Wei and Liu, Peter J},
  journal={Journal of machine learning research},
  volume={21},
  number={140},
  pages={1--67},
  year={2020}
}

@inproceedings{influence,
  title={Understanding black-box predictions via influence functions},
  author={Koh, Pang Wei and Liang, Percy},
  booktitle={International conference on machine learning},
  pages={1885--1894},
  year={2017},
  organization={PMLR}
}

@article{mislabeled,
  title={Estimating training data influence by tracing gradient descent},
  author={Pruthi, Garima and Liu, Frederick and Kale, Satyen and Sundararajan, Mukund},
  journal={Advances in Neural Information Processing Systems},
  volume={33},
  pages={19920--19930},
  year={2020}
}

@article{memorize,
  title={What neural networks memorize and why: Discovering the long tail via influence estimation},
  author={Feldman, Vitaly and Zhang, Chiyuan},
  journal={Advances in Neural Information Processing Systems},
  volume={33},
  pages={2881--2891},
  year={2020}
}

@article{interpretability,
  title={Post-hoc interpretability for neural nlp: A survey},
  author={Madsen, Andreas and Reddy, Siva and Chandar, Sarath},
  journal={ACM Computing Surveys},
  volume={55},
  number={8},
  pages={1--42},
  year={2022},
  publisher={ACM New York, NY}
}

@article{less,
  title={Less: Selecting influential data for targeted instruction tuning},
  author={Xia, Mengzhou and Malladi, Sadhika and Gururangan, Suchin and Arora, Sanjeev and Chen, Danqi},
  journal={arXiv preprint arXiv:2402.04333},
  year={2024}
}

@article{tinyllama,
  title={Tinyllama: An open-source small language model},
  author={Zhang, Peiyuan and Zeng, Guangtao and Wang, Tianduo and Lu, Wei},
  journal={arXiv preprint arXiv:2401.02385},
  year={2024}
}

@article{DS1000,
  title={DS-1000: A Natural and Reliable Benchmark for Data Science Code Generation},
  author={Yuhang Lai and Chengxi Li and Yiming Wang and Tianyi Zhang and Ruiqi Zhong and Luke Zettlemoyer and Scott Wen-tau Yih and Daniel Fried and Sida Wang and Tao Yu},
  journal={ArXiv},
  year={2022},
  volume={abs/2211.11501}
}

@article{mbpp,
  title={Program Synthesis with Large Language Models},
  author={Austin, Jacob and Odena, Augustus and Nye, Maxwell and Bosma, Maarten and Michalewski, Henryk and Dohan, David and Jiang, Ellen and Cai, Carrie and Terry, Michael and Le, Quoc and others},
  journal={arXiv preprint arXiv:2108.07732},
  year={2021}
}

@article{crosscodeeval,
  title={Mapping language to code in programmatic context},
  author={Iyer, Srinivasan and Konstas, Ioannis and Cheung, Alvin and Zettlemoyer, Luke},
  journal={arXiv preprint arXiv:1808.09588},
  year={2018}
}

@article{birdsql,
  title={Can llm already serve as a database interface? a big bench for large-scale database grounded text-to-sqls},
  author={Li, Jinyang and Hui, Binyuan and Qu, Ge and Yang, Jiaxi and Li, Binhua and Li, Bowen and Wang, Bailin and Qin, Bowen and Geng, Ruiying and Huo, Nan and others},
  journal={Advances in Neural Information Processing Systems},
  volume={36},
  year={2024}
}

@article{roberta,
  title={Roberta: A robustly optimized bert pretraining approach},
  author={Liu, Yinhan},
  journal={arXiv preprint arXiv:1907.11692},
  year={2019}
}

@article{agentless,
  title={Agentless: Demystifying llm-based software engineering agents},
  author={Xia, Chunqiu Steven and Deng, Yinlin and Dunn, Soren and Zhang, Lingming},
  journal={arXiv preprint arXiv:2407.01489},
  year={2024}
}

@article{spearman,
  title={Spearman rank correlation: overview},
  author={Zar, Jerrold H},
  journal={Wiley StatsRef: Statistics Reference Online},
  year={2014},
  publisher={Wiley Online Library}
}

@article{codeagent1,
  title={Communicative agents for software development},
  author={Qian, Chen and Cong, Xin and Yang, Cheng and Chen, Weize and Su, Yusheng and Xu, Juyuan and Liu, Zhiyuan and Sun, Maosong},
  journal={arXiv preprint arXiv:2307.07924},
  year={2023}
}

@article{codeagent2,
  title={Repocoder: Repository-level code completion through iterative retrieval and generation},
  author={Zhang, Fengji and Chen, Bei and Zhang, Yue and Liu, Jin and Zan, Daoguang and Mao, Yi and Lou, Jian-Guang and Chen, Weizhu},
  journal={arXiv preprint arXiv:2303.12570},
  year={2023}
}

@article{codesum1,
  title={Automatic code summarization via chatgpt: How far are we?},
  author={Sun, Weisong and Fang, Chunrong and You, Yudu and Miao, Yun and Liu, Yi and Li, Yuekang and Deng, Gelei and Huang, Shenghan and Chen, Yuchen and Zhang, Quanjun and others},
  journal={arXiv preprint arXiv:2305.12865},
  year={2023}
}

@article{codesum2,
  title={Code structure--guided transformer for source code summarization},
  author={Gao, Shuzheng and Gao, Cuiyun and He, Yulan and Zeng, Jichuan and Nie, Lunyiu and Xia, Xin and Lyu, Michael},
  journal={ACM Transactions on Software Engineering and Methodology},
  volume={32},
  number={1},
  pages={1--32},
  year={2023},
  publisher={ACM New York, NY}
}

@article{influence2,
  title={Do Influence Functions Work on Large Language Models?},
  author={Li, Zhe and Zhao, Wei and Li, Yige and Sun, Jun},
  journal={arXiv preprint arXiv:2409.19998},
  year={2024}
}

@article{bigscience,
  title={The bigscience roots corpus: A 1.6 tb composite multilingual dataset},
  author={Lauren{\c{c}}on, Hugo and Saulnier, Lucile and Wang, Thomas and Akiki, Christopher and Villanova del Moral, Albert and Le Scao, Teven and Von Werra, Leandro and Mou, Chenghao and Gonz{\'a}lez Ponferrada, Eduardo and Nguyen, Huu and others},
  journal={Advances in Neural Information Processing Systems},
  volume={35},
  pages={31809--31826},
  year={2022}
}

@misc{codescripts,
  title        = {{An Empirical Study on Data Influence-Based Pretraining Data Selection for Code Large Language Models}},
  howpublished = {\url{https://github.com/ZZR0/DIScore}},
  note         = {Accessed: 2025-10-23.}
}

@inproceedings{DataMixing,
  author       = {Jiasheng Ye and
                  Peiju Liu and
                  Tianxiang Sun and
                  Jun Zhan and
                  Yunhua Zhou and
                  Xipeng Qiu},
  title        = {Data Mixing Laws: Optimizing Data Mixtures by Predicting Language
                  Modeling Performance},
  booktitle    = {The Thirteenth International Conference on Learning Representations,
                  {ICLR} 2025, Singapore, April 24-28, 2025},
  publisher    = {OpenReview.net},
  year         = {2025},
  url          = {https://openreview.net/forum?id=jjCB27TMK3},
  bibsource    = {dblp computer science bibliography, https://dblp.org}
}

\end{document}